\documentclass{caosp302}
\usepackage{graphicx}

\articleNo{123}
\pubyear{2005}
\volume{35}
\volnumber{3}
\firstpage{1}
\received{May 1, 2007}
\accepted{August 28, 2007}

\begin{document}

%%%%%%%%%%%%%%%%%%%%%%%%%%%%%%%%%%%%%%%%%%%%%%%%%%%%%%%%%%%%%%%%%%%%%%%%%%%%%
%              R U N N I N G   P A G E   H E A D I N G S                     
% Odd page headings (except for the title page) are produced automatically
% and contain the title. If, and only if, the title of your article is too
% long the running head is ommitted in the printout; you can make your own
% running title by using the \htitle command, putting the shortened title
% between the curly brackets. \htitle should also be used when the
% subtitle is present: \htitle offers you a way how to include it into the
% headings. If you wish to see how it works simply remove the % sign from
% the beginnig of that line.
%
% Unlike the \htitle command, the \hauthor command is compulsory. It is
% used to produce even page headings and contains the names of the authors
% of an article.  All authors must be listed here, if possible. When
% authors' list is too long, you can abbreviate it by using "{\it et
% al.}". Authors' names are given in the form: initial(s) of the author's
% first name and surname. Authors are separated by a "," (comma) sign and
% the last one by "and".
%%%%%%%%%%%%%%%%%%%%%%%%%%%%%%%%%%%%%%%%%%%%%%%%%%%%%%%%%%%%%%%%%%%%%%%%%%%%%
%\htitle{A note to comet ejection process ...}
\hauthor{C.\,Schr\"oder, S.\,Hubrig and J.H.M.M.\,Schmitt}
%\hauthor{L.\,Neslu\v{s}an {\it et al.}}

%%%%%%%%%%%%%%%%%%%%%%%%%%%%%%%%%%%%%%%%%%%%%%%%%%%%%%%%%%%%%%%%%%%%%%%%%%%%%
%                       T I T L E                                            
% Capital letters in the title are only used at the beginning of the
% names. Don`t end the title by a "." (dot)
%%%%%%%%%%%%%%%%%%%%%%%%%%%%%%%%%%%%%%%%%%%%%%%%%%%%%%%%%%%%%%%%%%%%%%%%%%%%%
\title{Magnetic fields in X-ray emitting A-type stars}

%%%%%%%%%%%%%%%%%%%%%%%%%%%%%%%%%%%%%%%%%%%%%%%%%%%%%%%%%%%%%%%%%%%%%%%%%%%%%
%                       S U B T I T L E                                      
% You can use the subtitle, with the command \subtitle similar to the
% \title command.
%%%%%%%%%%%%%%%%%%%%%%%%%%%%%%%%%%%%%%%%%%%%%%%%%%%%%%%%%%%%%%%%%%%%%%%%%%%%%

%%%%%%%%%%%%%%%%%%%%%%%%%%%%%%%%%%%%%%%%%%%%%%%%%%%%%%%%%%%%%%%%%%%%%%%%%%%%%
%                   A U T H O R  N A M E S                                   
% Authors' names are separated by the \and commmand and their institutes
% are assigned by the \inst{n} command.
%
% When the name contains "Slovak" letters L,d,t,l with a caron, use an
% \additional softl, etc. command (examples given in the last table of
% this document) to produce typographically correct accented characters.
%%%%%%%%%%%%%%%%%%%%%%%%%%%%%%%%%%%%%%%%%%%%%%%%%%%%%%%%%%%%%%%%%%%%%%%%%%%%%
\author{
        C.\,Schr\"oder \inst{1,} 
      \and 
        S.\,Hubrig \inst{2}
      \and 
        J.H.M.M.\,Schmitt \inst{1}
       }

\institute{Hamburger Sternwarte, Gojenbergsweg 112, 21029 Hamburg, Germany\\
         \and 
           ESO, Casilla 19001, Santiago 19, Chile}

%\date{November 1, 2007}

\maketitle

\begin{abstract}
A common explanation for the observed X-ray emission of A-type stars is the presence of a hidden late-type companion.
While this hypothesis can be shown to be correct in some cases, there is also evidence suggesting that
low-mass companions cannot be the proper cause for the observed X-ray activity in all cases.
Babel \& Montmerle (1997) presented a theoretical framework to explain the X-ray emission for magnetic Ap/Bp stars, focusing on the A0p star IQ Aur.
We test if this theoretical model is capable to explain the observed X-ray emissions.
We present observations of 13 A-type stars that have been associated with X-ray emission detected by ROSAT.
To determine the mean longitudinal magnetic field strength we measured the circular polarization in the wings of the Balmer lines using FORS 1.
Although the emission of those objects with magnetic fields fits the prediction of the Babel \& Montmerle model,
not all X-ray detections are related to the presence of a magnetic field.
Additionally, the strengths of magnetic fields do not correlate with the X-ray luminosity
and thus the magnetically confined wind shock model cannot explain the X-ray emission from all investigated stars.
\end{abstract}

\section{Observations}
The observations have been carried out on August 28th 2006 with FORS 1 at the VLT Kueyen.
This multi-mode instrument is equipped with polarization analyzing optics comprising super-achromatic
half-wave and quarter-wave phase retarder plates, and a Wollaston prism with a beam divergence of 22\arcsec{} in standard resolution mode.
The grism 600B and the grism 1200B were used, which cover all H Balmer lines from H$\beta$ to the Balmer jump.
Most stars have been observed with the grism 600B at a spectral resolution of 2000.
%and three stars
%with the grism 1200B at a resolving power of R $\sim$4000 and a blue limit at 3885\,\AA{}.
Since we had only one observing night, we decided to observe only the two most promising targets
with the grism 1200B at a resolving power of R$\sim$4000 and a blue limit at 3885\,\AA{}.
A more detailed description of this technique was given by Hubrig et al.\ (2004a, 2004b).

\section{Result}
Out of 13 stars, seven are likely weakly magnetic.
Magnetic fields in HD\,147084, HD\,148898 and HD\,159312 are detected at a 3$\sigma$ level (see Fig.~1),
while for HD\,174240 and HD\,224392 they are detected at a 2$\sigma$ level.
For HD\,186219 and HD\,217186 the detection has been achieved just below the 2$\sigma$ level.
The measurements for the five stars HD\,163336, HD\,172555, HD\,186219, HD\,217186 and HD\,224361
yielded no detection, but a close inspection revealed Zeeman features in several lines in the Stokes V spectra.
These stars are therefore promising targets for further observations.
Only HD\,159217 showed no sign of a magnetic field.
We found no correlation between the X-ray luminosity and the measured magnetic field strength.
We have to note that because of the strong dependence of the longitudinal field on the rotational aspect,
its usefulness to characterize actual field strength distributions is rather limited (Hubrig et al.\ 2007).
This can be overcome by additional future observations to sample various rotation phases.
On the other hand, those stars with a detected magnetic field possess X-ray emission which fits the predicted values from the model by Babel \& Montmerle.

\begin{figure}
\centerline{\includegraphics[width=6.2cm,height=12.cm,angle=90,clip=,keepaspectratio=false]{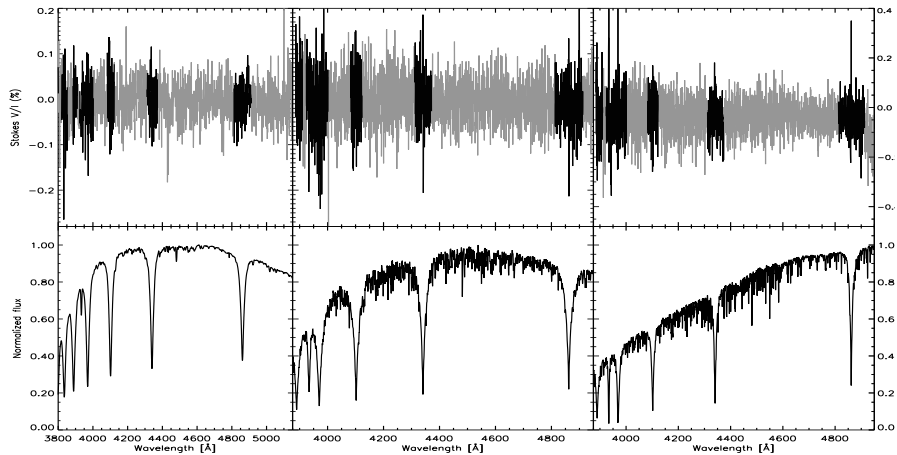}}
\caption{
Stokes V/I and normalized spectra of HD\,159312, HD\,148898 and HD\,147084, which have magnetic fields detected at a 3$\sigma$ level.
The black areas in the upper row indicate the regions used for determination of the magnetic fields from Balmer lines.
}
\end{figure}


\begin{thebibliography}{}
\bibitem{} Babel, J. \& Montmerle, T. 1997, {A\&A}, {323}, {121}
%\bibitem{} Hubrig, S., Bagnulo, S., Kurtz, D.W. et al. 2003, in ASP Conference Series, ed. L. A. Balona, H. F. Henrichs \& R. Melupe, {114+}
\bibitem{} Hubrig, S., Kurtz, D.W., Bagnulo, S., et al. 2004a, {A\&A}, {415}, {661}
\bibitem{} Hubrig, S., Szeifert, T., Sch\"oller, M. et al. 2004b, {A\&A}, {415}, {685}
\bibitem{} Hubrig, S., North, P., Sch\"oller, M. \& Mathys, G. 2007, {Astronomische Nachrichten}, {328}, {475}
\end{thebibliography}
\end{document}